# On the Virtues of Automated QSAR – The New Kid on the Block

Marcelo T. de Oliveira*[1,2], Edson Katekawa[2]

1 Instituto de Química de São Carlos, Universidade de São Paulo, Avenida Trabalhador são-carlense 400, São Carlos, SP 13566-590, Brazil

2 Instituto de Física de São Carlos, Universidade de São Paulo, Brazil

*Author for correspondence: mtavareso@usp.br

Quantitative Structure-Activity Relationship (QSAR) has proved an invaluable tool in medicinal chemistry. Data availability at unprecedented levels through various databases have collaborated to a resurgence in the interest for QSAR. In this context, rapid generation of quality predictive models is highly desirable for hit identification and lead optimization. We showcase the application of an automated QSAR approach, which randomly selects multiple training/test sets and utilizes machine-learning algorithms to generate predictive models. Results demonstrate that AutoQSAR produces models of improved or similar quality to those generated by practitioners in the field but in just a fraction of the time. Despite the potential of the concept to the benefit of the community, the AutoQSAR opportunity has been largely undervalued.

**Background**

Hansch's seminal publication on the correlation of partition coefficients and biological activity evaluating the effect of substituents through the application of straightforward regression methods laid the foundations of the field we recognize as Quantitative Structure-Activity Relationship (QSAR) [1-2]. Over half a century later and thousands of papers disclosing applications in numerous areas including medicinal chemistry, [3-5] environmental, [6-9] food, [10-12] and material [13-16] sciences, as well as regulatory [17-19] across academy, industry, and government, QSAR has proved an invaluable tool for experts and non-experts alike in identifying patterns in datasets and assisting in the process of making quantitative predictions [20-21]. In drug discovery, QSAR represents a ligand-based method, which generates models by building relationships between structural information or properties for a set of compounds and related experimental data of interest (e.g., biological activity) as the target property [22-23]. These models have been particularly useful to lead optimization [24].

As in any field, QSAR has experienced highs and lows through its history. In the context of well-established structure-based methods such as molecular docking enjoying the spotlight, a major review presented the provocative question 'QSAR Modeling: Where have you been?', as part of the title [25]. From another downside perspective, there are skeptical critics who deem QSAR models as plain black boxes carrying no ability to promote creative drug design [26]. Nonetheless, as we witness the recent explosion of medicinal chemistry information filling databases of public domain in the era of 'big data', one can easily understand the current resurgence in the interest for QSAR after years of dominance of structure-based approaches [27].

Developments in QSAR have been constant since its early days. Despite the abundance of data becoming available and easy access to sizeable collections for prospective studies, practitioners have been creating models essentially in the same manner as 30 or more years ago. After curating a learning set of interest, a typical QSAR routine starts with the computation of descriptors followed by splitting the dataset into training and test sets, which then allows construction of QSAR models; and as a final step, validation of generated models. As access to large datasets has become commonplace and descriptors are usually readily computed, one can turn attention to the next couple steps, which represent bottlenecks in the process. Considering the many variables brought into play, manual selection of best training and test sets is a laborious process as a myriad of configurations are possible. Additionally, QSAR models can be created following a variety of methods leading to a vast number of possible results, which would further stretch the workload. Practitioners would largely benefit from automation of both tasks resulting in a massive reduction of the time necessary to achieve end-result models following random selection of training/test sets and various built-in algorithms to generate a range of models.

Automated QSAR has been in the making for quite a few years. While searching for QSAR as a keyword returns over 15,000 hits (articles and reviews, only) at Web of Science (Figure 1), AutoQSAR or "automated QSAR" produces just a handful entries (as late June, 2017) [28-31]. Early references to preliminary automated QSAR efforts appeared in the 2000s [32-37]. A major contribution arrived in 2011 when an AstraZeneca's team reported their experience related to the continuous update of QSAR models for in-house ever-growing compound collections [31].

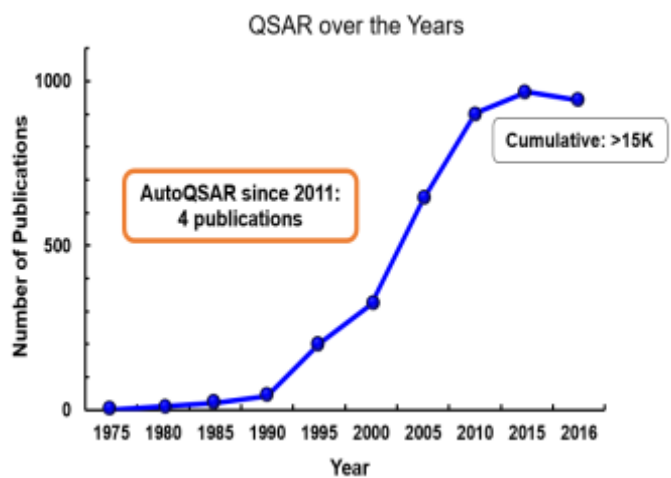

**Figure 1.** QSAR and AutoQSAR publications over the years.

To date, however, AutoQSAR has attracted only scant attention from the community. In contrast, there are quite a few applications currently available running automated QSAR protocols including (but not only): AstraZeneca's AutoQSAR [29], AutoQSAR [38,39], Auto-Modeller [40], QSAR workbench [31], and VLifeAutoQSAR [41]. Among web-based services, we mention OCHEM [42] database and ChemModLab [43]. They all offer a rapid and simplified QSAR solution just a few clicks away to a range of models skipping iterative steps that require some statistical dexterity to the benefit especially of non-experts. AutoQSAR applications basically differ on descriptors automatically computed and methods (e.g., decision trees, PLS, random forests) available to generate models.

For this exercise, we applied an AutoQSAR module introduced to Schrödinger's Small-Molecule Drug Discovery suite early last year [38]. Instead of considering a single model derived from a single training set, AutoQSAR randomly selects combinations of training and test sets (75% and 25%, respectively, on default mode) distributing data following the same pattern for the learning set. Then, it computes 497 physicochemical and topological descriptors as well as 2D fingerprints (dendritic, linear, MOLPRINT2D, or radial) through the Canvas module. These are used in combination with machine-learning methods (Bayes, KPLS, MLR, PCR, PLS, and RP) available to build various predictive models. Among these, only Bayes and KPLS apply both descriptors and fingerprints. To rank models on the basis of merit, which is critical to any AutoQSAR scheme, the module offers a robust score function that evaluates each model in respect to its accuracy in relation to training and test sets. That is, $r^2$ and $q^2$, respectively, are taken into account (Figure 2).

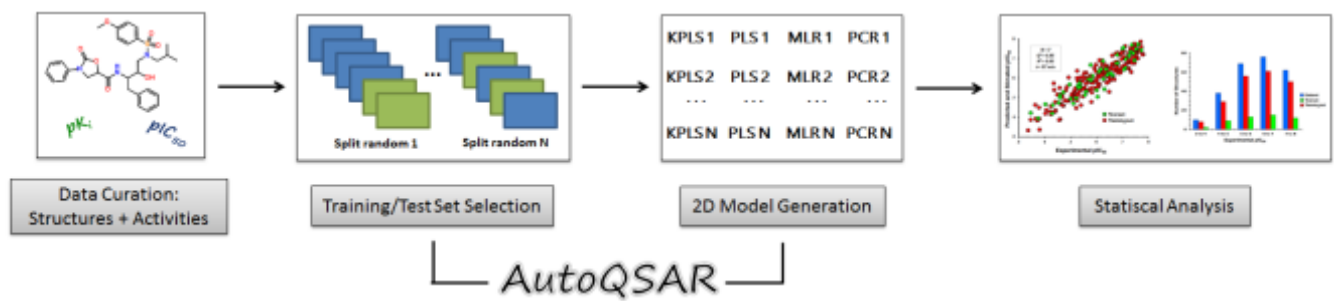

**Figure 2.** QSAR workflow highlighting AutoQSAR steps of selecting training/test sets and model generation.

**Automated QSAR in action**

To build the case of AutoQSAR, we set out to generate QSAR predictive models from published datasets. We aimed to compare results generated in our group using AutoQSAR (unpublished data from the authors) to models available in the literature produced using traditional approaches. Dixon and co-workers [39] recently reported the analysis of AutoQSAR models comparing to their previously published results using machine-learning methods [44]. Herein, we analyse AutoQSAR results in relation to QSAR models from various research groups. We were particularly interested in confronting the quality of both models and evaluate the time required using the AutoQSAR approach. Towards these goals, we chose 7 examples illustrating typical applications of QSAR in the early stages of drug discovery when searching for a hit or lead compound. Such cases represent the majority of QSAR publications in the medicinal chemistry literature. The selection includes from a single series of compounds to a few classes for the same target. Learning sets varied from 35 to 255 compounds for a defined biochemical target (in parenthesis) related to the following diseases: Alzheimer's (*h*BACE1) [45], asthma (CRTh2) [46], cardiovascular (ABHD6) [47], Chagas' (cruzain) [48], diabetes (PTP1B) [49], malaria (*Pf*DHODH) [50], and schistosomiasis (*Sm*TGR) [51]. All examples represent recent and validated models. It is noteworthy that half examples initially selected had an issue (e.g., clarity, duplicates, typographic mistakes) that ruled out its use for the present study.

AutoQSAR protocol was applied to the datasets above to select test and training sets. Models were produced at training sets 70 to 80% of the dataset at 1% intervals; and 99 models were built at each interval. Maximum pair correlation was set to 0.99 while other parameters were kept as default. In selecting best AutoQSAR model from the many produced, we used the highest score as main criteria followed by $Q^2$ ($R^2$ for the test set). $Q^2$ values were compared to $R^2_{pred}$ from literature examples. Most AutoQSAR models (5/7) showed a substantial improvement in the predictive power. For instance, asthma's CRTh2 original model produced $R^2_{pred}$ value of 0.64 while AutoQSAR $Q^2$ value was 0.88 (Figure 3). Models related to a couple of learning sets (Chagas' and HIV) had similar predictive capacity. AutoQSAR underperformed in comparison to the original model and only slightly in just one case (schistosomiasis' *Sm*TGR), $R^2_{pred}$ 0.96 versus $Q^2$ 0.90. Nonetheless, we obtained a much simpler AutoQSAR solution with fewer components (*N*=2) than in the original model (*N*=6) [28]. Interestingly, all best models obtained here were invariably a KPLS-related solution. The main advantage of AutoQSAR methods, however, is the time necessary to generate a collection of models even on a basic consumer-grade desktop as in our case. Times provided represent wall times, which varied according to the size of the learning set from 18 to 97 min; and 40 min as average (Figure 3). Even malaria (*Pf*DHODH) as the largest dataset had calculations for all models completed in less than a day (*ca.* 19 h). That represents a dramatic reduction when compared to the many days or weeks on end of worktime to generate a single or just a few models for a learning set.

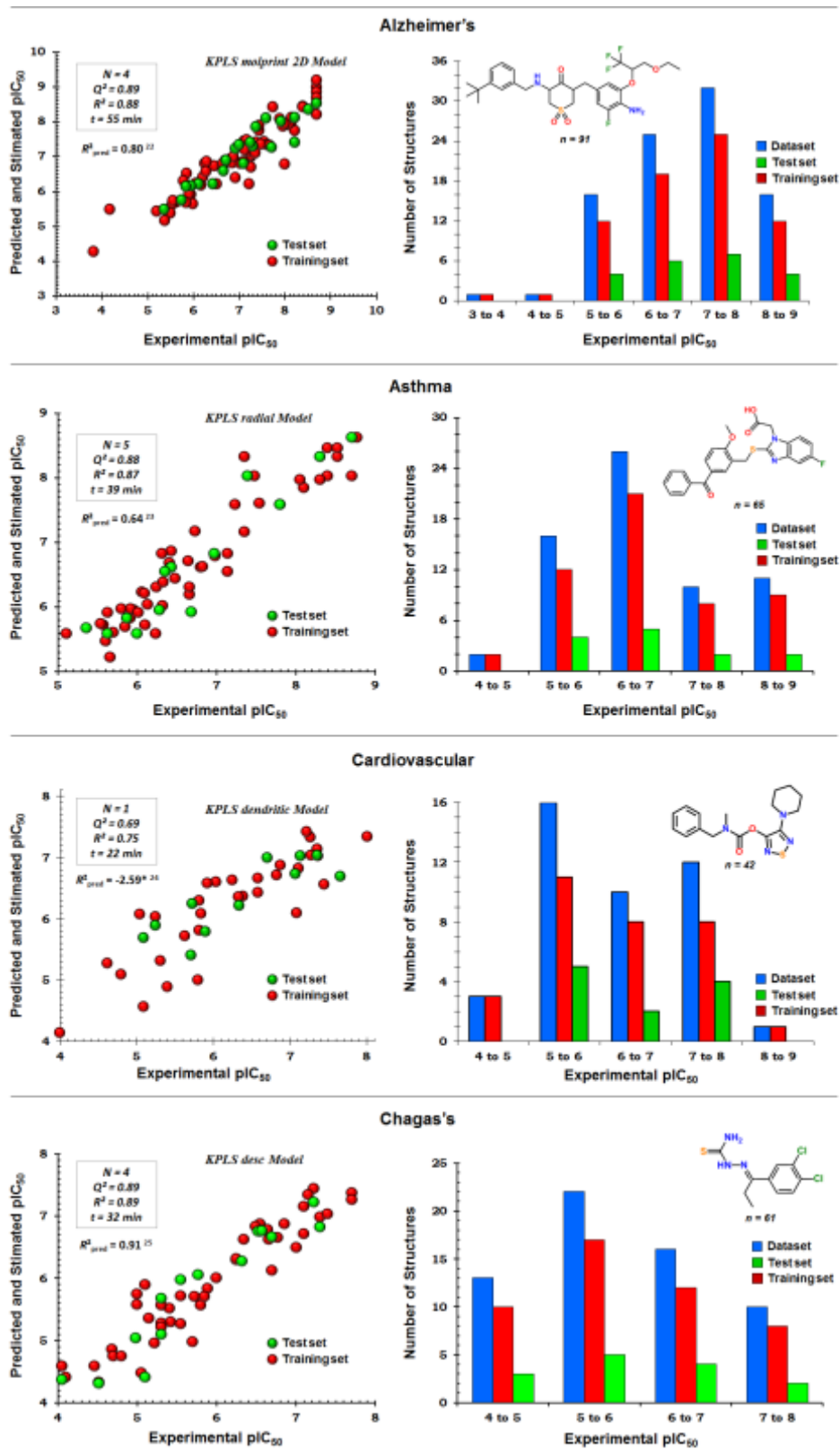

**Figure 3.** Predicted and simulated pIC$_{50}$ versus experimental pIC$_{50}$ for best generated model. Training and test sets are identified. Values of $Q^2$, $R^2$, $R^2_{pred}$, and times to generate models are also provided. Note that all best models were KPLS (left). Training and test set distribution after AutoQSAR as well as experimental dataset distribution (right). *$R^2_{pred}$ was unavailable for cardiovascular (ABHD6); value provided was calculated from the original reference.

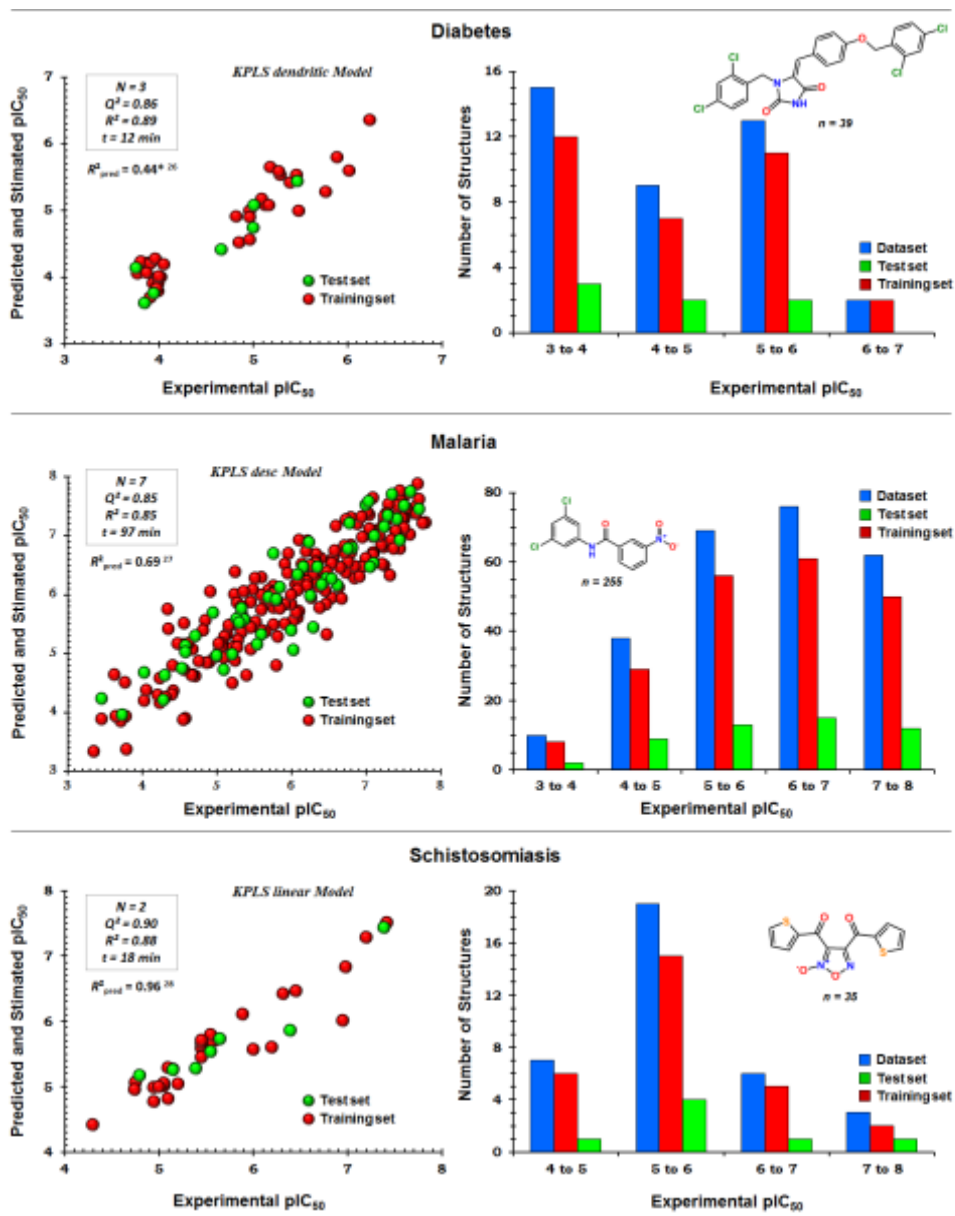

**Figure 3 (continued).** Predicted and simulated pIC$_{50}$ versus experimental pIC$_{50}$ for best generated model. Training and test sets are identified. Values of $Q^2$, $R^2$, $R^2_{pred}$, and times to generate models are also provided. Note that all best models were KPLS (left). Training and test set distribution after AutoQSAR as well as experimental dataset distribution (right). *$R^2_{pred}$ was unavailable for diabetes (PTP1B); value provided was calculated from the original reference.

**Conclusion & Future Perspectives**

Traditionally, creation of high-quality QSAR models has required significant expertise and being labor intensive. Following best practices, AutoQSAR approaches democratizes creation and application of QSAR models through automation. Bearing in mind these clear advantages, the obvious question about the reasons why automated QSAR protocols have not been more adopted beyond the handful examples aforementioned remains unanswered. Perhaps AutoQSAR is too recent and a great deal of the community have not had the opportunity to try it out for themselves; or even that is it unheard of by many as we have noticed among colleagues. Cumming and co-workers [52], in 2013, went further to offer the bold argument to which we agree that the community lacks confidence in the capacity of machines to produce models – similar or better than those built by experts in the field. Therefore, implying it is a case of a psychological barrier to be overcome. Beyond controversies and differences of opinion regarding QSAR, we offer our AutoQSAR experience based on the bit of 'taste' we have had for the past few months. For its considerable benefits, it is clear to us that it is a powerful tool. If only appreciated, automated QSAR protocols have huge potential for various applications considering speeding up quality model generation, thus facilitating tackling with confidence increasingly larger datasets by QSAR experts and non-experts alike.

**Financial & competing interests disclosure**

This work was supported by the State of São Paulo Research Foundation (FAPESP, Fundação de Amparo à Pesquisa do Estado de São Paulo), grant 2013/07600-3. E.K. acknowledges financial support provided by FAPESP, grant 2014/50623-7. M.T.O acknowledges the Coordination for the Improvement of Higher Education Personnel (CAPES, Coordenação de Aperfeiçoamento de Pessoal de Nível Superior) for research grant. E.K. is an employee at Chemyunion Ltda, Sorocaba, São Paulo, Brazil. The authors have no other affiliations or financial involvement with any organization or entity with a financial interest or financial conflict with the subject matter or materials discussed in the manuscript from those disclosed.

No writing assistance was utilized in the production of this manuscript.

# Executive summary

## QSAR in medicinal chemistry

- QSAR has long been used in ligand-based drug discovery efforts to generate predictive models to assist in the identification of hit compounds and lead optimization.
- Through the decades, QSAR has faced criticism for inadequacies in assisting in drug design.
- There has been a resurgence in interest in QSAR due to data availability at unprecedented levels.

## The Automated QSAR opportunity

- Model generation using QSAR has been done essentially in the same manner for the past 30 years or so.
- In the era of 'big data', automation of tasks enabling data crunching to identify patterns is highly desirable. Automation efforts in QSAR has been described since 2000s.
- Automated QSAR workflows assist in model generation by identifying training and test sets as well as producing various models using machine-learning methods.
- As shown herein, AutoQSAR produces models of similar or improved quality to those generated by experts using traditional approaches but in a fraction of the time.

## Conclusion

- Considering the limited number of publications, we argue that not many users are aware of the automated QSAR opportunity or have tried it for themselves. Scepticism that machines could outperform practitioners may also explain this trend.
- Automated QSAR protocols would be particularly useful to tackle large datasets or systems which requires construction of various models in a reduced amount of time.
- Ultimately, automated QSAR democratizes generation and application of QSAR models following best practices in the field.